\documentclass[]{aa}  
\usepackage{graphicx}
\usepackage{natbib}
\usepackage{subfig}
\usepackage{epsfig} 
\usepackage{ifpdf} 
\usepackage{fancyvrb}
\usepackage{txfonts}
\usepackage{xcolor}
\usepackage{soul}
\usepackage{mathtools}

\begin{document} 

\title{Fermi acceleration along the orbit of $\eta$ Carinae}

\author{M. Balbo \and R. Walter}
          
\institute{Astronomy Department, University of Geneva, Chemin d’Ecogia 16, 1290 Versoix, Switzerland
}

\date{Received September 2, 2016; accepted April 20, 2017}
% \abstract{}{}{}{}{} 
% 5 {} token are mandatory
 
\abstract
   {The $\eta$ Carinae binary system hosts one of the most massive stars, which features the highest known mass-loss rate. This dense wind encounters the much faster wind expelled by the stellar companion, dissipating mechanical energy in the shock, where particles can be accelerated up to relativistic energies and subsequently produce very-high-energy $\gamma$-rays.}
   {We aim at comparing the variability of the $\gamma$-ray emission of $\eta$ Carinae along the binary orbit with the predictions of simulations to establish the nature of the emission and of the seed particles.}
   {We have used data from the Fermi Large Area Telescope obtained during its first seven years of observations and spanning two passages of $\eta$ Carinae at periastron. We performed the analysis using the new PASS8 pipeline and its improved instrument response function, extracting low and high-energy light curves as well as spectra in different orbital phase bins. We also introduced particle acceleration in hydrodynamic simulations of the system, assuming a dipolar magnetic field generated by the most massive star, and compared the $\gamma$-ray observations with the predictions of diffuse shock acceleration in a multi-cell geometry.}
   {The main source of the $\gamma$-ray emission originates from a position compatible with $\eta$ Carinae and located within the Homunculus Nebula. Two emission components can be distinguished. The low-energy component cuts off below 10 GeV and its flux, modulated by the orbital motion, varies by a factor less than 2. Short-term variability occurs at periastron. The flux of the high-energy component varies by a factor 3-4 but is different during the two periastrons. The variabilities observed at low energy, including some details of them, and those observed at high energy during the first half of the observations, match the prediction of the simulation, assuming a surface magnetic field of 500 G. The high-energy component and the thermal X-ray emission were weaker than expected around the second periastron suggesting a modification of the wind density in the inner wind collision zone.}
   {Diffuse shock acceleration in the complex geometry of the wind collision zone of $\eta$ Carinae provides a convincing match to the observations and new diagnostic tools to probe the geometry and energetics of the system. This demonstrates that Fermi acceleration is at work in the wind collisions and that a few percent of the shock mechanical energy are converted into particle acceleration. Further observations are required to understand the periastron-to-periastron variability of the high-energy component and to associate it firmly with hadronic origin. We estimate that $\eta$ Carinae is a pevatron at periastron and is bright enough to be detected by IceCube after many years of observations. Orbital modulations of the high-energy component can be distinguished from those of photo absorption by the four large size telescopes of the Cherenkov Telescope Array to be placed in the southern hemisphere.}

\maketitle

\section{Introduction}

$\eta$ Carinae is the most luminous massive binary system of our galaxy and the first one to have been detected at very high energies without hosting a compact object\footnote{So far this list also counts the interesting case of WR11~\citep{2016MNRAS.457L..99P}}. It is composed by one of the most massive stars known ($\eta$ Car A) with an initial mass estimated above M$_{\rm{A}} \gtrsim 90 M_{\odot}$ \citep{2001ApJ...553..837H} and of a companion ($\eta$ Car B) believed to be an O supergiant or a WR star. 

$\eta$ Carinae has been studied across the whole electromagnetic spectrum from the radio \citep{2007MNRAS.382..382M} to the TeV \citep{2012MNRAS.424..128H}, passing through the infrared \citep{1994MNRAS.270..364W,2004MNRAS.352..447W}, optical \citep{1996ApJ...460L..49D,2000ApJ...528L.101D,2008MNRAS.384.1649D}, X-rays \citep{2001ApJ...547.1034C,2008MNRAS.388L..39O,2008A&A...477L..29L,2010A&A...524A..59L}, and $\gamma$-rays \citep{2009ApJ...698L.142T,2010ApJ...723..649A,2011A&A...526A..57F}. The presence of the companion star so far has been only indirectly inferred from the effect of the wind-wind collision (in particular by the X-ray emission from a multi keV plasma) and from the variable ultraviolet emission photoionizing nearby circumstellar clouds \citep{2005ApJ...633L..37I,2010ApJ...710..729M}.

$\eta$ Car A is accelerating a very dense wind with a mass-loss rate of $\sim 8.5\times 10^{-4}$ M$_{\odot}$ yr$^{-1}$ and a terminal wind velocity of $\sim 420$ km s$^{-1}$ \citep{2012MNRAS.423.1623G}. Its companion probably emits a fast low-density wind at $10^{-5}$ M$_{\odot}$ yr$^{-1}$ reaching a velocity of 3000 km s$^{-1}$ \citep{2002A&A...383..636P,2005ApJ...624..973V,2009MNRAS.394.1758P}. 

The regular modulation detected in the X-ray light curves suggests that the two stars are located in a very eccentric orbit \citep{2001ApJ...547.1034C,2008MNRAS.388L..39O}. The estimated orbital period at the epoch of the Great Eruption that happened between 1837-1856 was $\sim5.1$ yr and since then has increased up to the current $\sim5.54$ yr \citep{2004MNRAS.352..447W,2005AJ....129.2018C,2008MNRAS.384.1649D} owing to the huge quantities of mass and energy dissipated during the past century. During the Great Eruption, $\eta$ Carinae experienced a huge outburst ejecting an impressive quantity of mass estimated as $10-40~M_\odot$ \citep{2010MNRAS.401L..48G} at an average speed of $\sim 650$ km s$^{-1}$ \citep{2003AJ....125.1458S}, giving rise to the formation of the Homunculus Nebula and becoming one of the brightest stars of the sky. The energy released in such a catastrophic event ($10^{49-50}$ erg) was comparable with a significant fraction of the energy emitted by a supernova explosion. 

Given the high eccentricity of the orbit, the relative separation of the two stars varies by a factor $\sim20$, reaching its minimum at periastron, when the two objects pass within a few AU of each other; the radius of the primary star is estimated as 0.5 AU. In these extreme conditions their supersonic winds interact forming a colliding wind region of hot shocked gas where charged particles can be accelerated via diffusive shock acceleration up to high energies \citep{1993ApJ...402..271E,2003A&A...409..217D,2006ApJ...644.1118R}. As these particles encounter conditions that vary with the orbital phase of the binary system, one can expect a similar dependency in the $\gamma$-ray emission.

The hard X-ray emission detected by INTEGRAL \citep{2008A&A...477L..29L} and Suzaku \citep{2008MNRAS.388L..39O}, with an average luminosity $(4$-$7)\times10^{33}$ erg s$^{-1}$, suggested the presence of relativistic particles in the system. The following year AGILE detected a variable source compatible with the position of $\eta$ Carinae \citep{2009ApJ...698L.142T}. Other $\gamma$-ray analyses followed, which reported a luminosity of $1.6\times10^{35}$ erg s$^{-1}$ \citep{2010ApJ...723..649A,2011A&A...526A..57F,2012A&A...544A..98R}, and suggested the presence of hard component in the spectrum around periastron, which subsequently disappeared around apastron. Such a component has been explained through $\pi^0$ decay of accelerated hadrons interacting with the dense stellar wind \citep{2011A&A...526A..57F}, or interpreted as a consequence of $\gamma$-ray absorption against an ad hoc distribution of soft X-ray photons \citep{2012A&A...544A..98R}.

An alternative acceleration scenario suggested by \cite{2010ApJ...718L.161O}, which associated particle acceleration to the blast wave of the 1843 Great Eruption and foresaw a constant flux emission, was ruled out by the variability detected in the Fermi LAT light curves.

At even higher energies, the observation by \cite[the][]{2012MNRAS.424..128H} did not lead to any significant detection, raising only an upper limit at energies $\gtrsim 500$ GeV. This in turn would imply a sudden drop in the $\gamma$-ray flux, which could be related to a cut-off in the accelerated particle distribution or to severe $\gamma-\gamma$ absorption.

Our study starts with a new analysis of the Fermi-LAT data, including 25\% more data than published previously and the latest version of the pipeline processing. We then present the results of a simulation of particle acceleration in the colliding wind, based on detailed hydrodynamic simulations of \cite{2011ApJ...726..105P}, and compare these with the observations. These comparisons are very successful and lead to a number of conclusions reinforcing our previous interpretations \citep{2008A&A...477L..29L,2011A&A...526A..57F}.

\section{Fermi-LAT data analysis}
Launched on 2008 June 11, the Large Area Telescope (LAT) on-board the Fermi Gamma-ray Space Telescope is the most sensitive $\gamma$-ray telescope to date, covering an energy range from 20 MeV to 300 GeV~\citep{2009ApJ...697.1071A}. The LAT is characterized by large field of view ($2.4$ sr at 1~GeV) and collecting area ($\sim6500$ cm$^2$ at 1~GeV), a low deadtime ($<100~\mu$s per event), a high time resolution ($<10~\mu$s), and an energy dependent point-spread function (PSF) improving from $\sim5^{\circ}$ ($68\%$ containment) at 100~MeV to $\sim0.1^{\circ}$ at 40~GeV. The LAT consists of a charged particle tracker, a calorimeter, and an anti-coincidence system.

The electron-positron pair conversion tracker is made of 36 layers of silicon strip detectors to track charged particles, interleaved with 16 layers of tungsten foil to facilitate the conversion of $\gamma$-rays to pairs, these layers are comprised of 12 thin layers in the front section followed by 4 thick layers in the back, of 0.03 and 0.18 radiation length, respectively. The calorimeter, located at the bottom of the instrument, has $\sim8.5$ total radiation lengths of caesium iodide to measure the total event energy. Given the intense background of charged particles from cosmic rays and trapped radiation at the orbit of the Fermi satellite, the instruments are protected by a segmented anti-coincidence detector used to reject charged-particle background events. More information about the LAT is provided in~\cite{2009ApJ...697.1071A}, the LAT in-flight calibration is described in ~\cite{2009APh....32..193A}, \cite{2012ApJS..203....4A}, and \cite {2012APh....35..346A}.

We performed the analysis of the recent Fermi-LAT data since the beginning of the regular survey-mode observations on 2008 August 4, until 2015 July 1, more precisely mission elapsed time (MET): 239557417 to 457485024 s.
We used only \textit{source} class data (i.e. reconstructed events with high probability of being photons) that have been reprocessed with the PASS8\footnote{http://fermi.gsfc.nasa.gov/ssc/data/analysis/documentation/\\/Pass8$\_$usage.html} pipeline and subsequently analysed using the Fermi Science Tool v10r0p5 package\footnote{Available on the Fermi Science Support Centre (FSSC) website: http://fermi.gsfc.nasa.gov/ssc/data/analysis/software/}. With respect to the previous PASS7, the new PASS8 introduced improvements in the reconstruction of the event direction, energy measurement, event selection, ghost handling, and track/anti-coincidence detector matching information. All these yield to an enhancement in the entire performance of the telescope. These enhancements include higher acceptance, larger field of view, smaller PSF, better energy resolution, deeper differential sensitivity, smaller systematic uncertainties, and the introduction of four subgroups of PSF and energy dispersion to improve even further the containment radius and energy resolution at the expense of the statistics. We selected data with \texttt{evclass=128} corresponding to the source class and \texttt{evtype=3} indicating that both categories of photons converted in the front and in the back part of the LAT were selected. We followed the analysis recommendations of the Fermi-LAT team for the time and event selection, rejecting photons with apparent zenith angle greater than $90^{\circ}$ in order to minimize the background due to the atmospheric $\gamma$-rays originated from the Earth's limb, which lies at a zenith angle of $\sim113^{\circ}$. We did not perform a zenith angle cut based on the region of interest (ROI) with \texttt{gtselect}, but did correct the exposure with the \texttt{zmax=90} option in the \texttt{gtltcube} tool. 

With its orbital period of 96.5 minutes, its maximal rocking angle of $60^{\circ}$ and its precession period of 53.4 days, Fermi-LAT spends more than $80\%$ of its operative time in survey mode providing a uniform coverage of the sky every two orbits ($\sim3$ hours). As a consequence, our target of interest is not always visible, so knowing the exact position and orientation of the satellite from the spacecraft files at each time, we ran \texttt{gtmktime} to obtain all the good time intervals in which to perform our analysis; we also excluded the readout dead-time $(\sim9\%)$ and the time intervals corresponding to the South Atlantic Anomaly ($\sim13\%$) when data taking is suspended. Finally data are filtered to accept only those events flagged with \texttt{(DATA\_QUAL==1) \&\& (LAT\_CONFIG==1)}, which exclude bad quality events and instrument configuration not recommended for scientific analysis, respectively.

Given the relative distance between the apparent position of the Sun during the year and the nominal position of $\eta$ Carinae, we can clearly neglect  the $\gamma$-ray contribution from our local star. The same is also valid for the Moon.

The instrument response function (IRF) of the Fermi-LAT, i.e. the description of the instrument performance provided for data analysis, strongly depends on the energy~\citep{2012ApJS..203....4A}; thus to better exploit its performance we made a separate analysis for photons above and below 10 GeV. This threshold was chosen because of the spectral shape of $\eta$ Carinae, as explained in the third paragraph of Sect. \ref{sec:low_energy}.

\subsection{High-energy emission: 10-300 GeV} \label{sec:high_energy}

For the high-energy analysis, we took into account only photons arriving within $3^{\circ}$ from the nominal position of $\eta$ Carinae (R.A.=161.264775, Dec=-59.684431), as the total (front+back) $95\%$ PSF containment angle above 10 GeV is smaller than $1^{\circ}$ on-axis. 
We created a sky model using the 3FGL Fermi-LAT four-year catalogue \citep{2015ApJS..218...23A} including all sources up to $1^{\circ}$ outside of the ROI. We used the same source spectral models as indicated in the catalogue, leaving the normalization parameter free to vary for all those sources within $2^{\circ}$ from $\eta$ Carinae and with an average test statistic\footnote{Source detection significance can be described by the likelihood test statistic value $TS=-2Log(L_{max,0}/L_{max,1})$, which compares the ratio of two values that are obtained by a maximum-likelihood procedure. \textit{$L_{max,0}$} is the likelihood for a model without an additional source at a specified location (the null-hypothesis), and \textit{$L_{max,1}$} is the maximum-likelihood value for a model including an additional source or one more free parameters.} (TS) reported from the catalogue greater than $100$ (corresponding to a detection significance $\sim 10\sigma$) or presenting a variability index higher than $72.44$, which indicates a $99\%$ confidence probability that the source is variable on a monthly timescale \citep{2015ApJS..218...23A}. The only exception to the model was made for the source \object{3FGL J1043.6-5930} (hereafter J1043), which does not satisfy the condition $TS>100$ but is only $0.27^{\circ}$ away from $\eta$ Carinae \citep[for a 68\% containment radius of $0.23^{\circ}$ above 10 GeV, ][]{2015ApJS..218...23A}. This source has been modelled in the 3FGL catalogue as a power law (PL) with index $\Gamma=2.07\pm0.11$ and we decided to leave its normalization parameter free as well.
For completeness we also checked if there was any source from the FAVA \citep{2013ApJ...771...57A} weekly flare list\footnote{http://fermi.gsfc.nasa.gov/ssc/data/access/lat/FAVA/index.php} present in our model. As the closest source lies more than $6^{\circ}$ away from the binary system, we did not take any particular precaution. Finally, with the only exception of $\eta$ Carinae, no other source is mentioned in the second Fermi-LAT catalogue of High-Energy Sources \citep{2016ApJS..222....5A}.

$\eta$ Carinae lies on the tangential projection of the Carina-Sagittarius Arm of the Milky Way, less than $1^{\circ}$ away from the Galactic plane. Consequently a correct description of the diffuse Galactic $\gamma$-ray background plays a key role in the analysis. To reproduce this emission we used the \texttt{gll\_iem\_v06.fits}\footnote{http://fermi.gsfc.nasa.gov/ssc/data/access/lat/BackgroundModels}~\citep{2016ApJS..223...26A} Galactic background model. Unfortunately, for several reasons the representation of the diffuse emission is a serious concern in the Carina region, becoming more and more delicate towards lower energies (see Sect. \ref{sec:low_energy}). As a matter of fact a significant number of sources from the catalogue \citep{2015ApJS..218...23A} are flagged with "c" in that region, indicating that they are considered to be potentially confused with the Galactic diffuse emission. In order to obtain a more realistic representation of the Galactic diffuse emission in such a small region we decided to let its normalization free. The isotropic background component has been represented using the model \texttt{iso\_P8R2\_SOURCE\_V6\_v06.txt$^{5}$}.

In the high-energy range we described the emission coming from $\eta$ Carinae (its signal to noise is very limited) with a simple PL model dN/dE = N$_{{\rm E}_{min}-{\rm E}_{max}}(\Gamma+1){\rm E^\Gamma/(E^{\Gamma+1}_{max}-E^{\Gamma+1}_{min})}$, where N$_{{\rm E}_{min}-{\rm E}_{max}}$ indicates the integrated photon flux over the fitting energy range $[{\rm E}_{min},{\rm E}_{max}]$ and $\Gamma$ represents the PL index.

Exploiting this model we performed a maximum-likelihood analysis \citep{1996ApJ...461..396M} on the previously described sample of photons, using the \texttt{P8R2\_SOURCE\_V6} IRF and the \texttt{gtlike} tool. Checking the results of the first fit, we found that the significance of some of the sources, left free in our model, were below $5\sigma$. This does not contradict the catalogue results because even though we are increasing the statistic covering a longer period with our data sample, we previously took the average significance over the whole energy band in our choice of the model, while here we are performing the analysis only above 10 GeV. Thus we proceeded to freeze those sources parameters to the catalogue values. We did the same for all the "c" flagged sources, as they might be just artefacts of a bad representation of the diffuse Galactic model at lower energies. Finally we let free the normalization parameter of the Galactic diffuse emission and isotropic background, and the fit resulted in an integrated flux for $\eta$ Carinae of F$_{\rm 10-300GeV}=(5.06\pm0.52) \times 10^{-10}$ ph cm$^{-2}$ s$^{-1}$ and index $\Gamma=-2.28 \pm 0.15$ with a TS value of 366 (significance $\gtrsim 19 \sigma$). Those results represent an average over the whole time period of nearly seven years and are in good agreement with that reported in \cite{2015A&A...577A.100R}. The slightly lower flux value could be explained by the lower than average flux in the data of the previous year \citep{2015AAS...22534415C}.

The resulting normalization factor of the Galactic diffuse emission differs significantly from unity. This is a known problem in the Carina region. Running again the fit, forcing the normalization of the Galactic diffuse emission component to unity, we obtain a $22\%$ higher integrated flux for $\eta$ Carinae, while the $\Gamma$ index remains unchanged. On the contrary, the isotropic background does not present specific issues on the Galactic plane, so we decided to freeze it to unity and let vary just the Galactic component normalization. Running again the likelihood, the fit yields a Galactic normalization of $1.95\pm0.06$, a slightly lower integrated flux for $\eta$ Carinae F$_{\rm 10-300GeV}=(5.00\pm0.51) \times 10^{-10}$ ph cm$^{-2}$ s$^{-1}$ and an index $\Gamma=-2.27 \pm 0.15$, for a significance of nearly $19\sigma$. Finally, as we do not expect any temporal variation of the Galactic diffuse emission we kept its normalization frozen to the seven-year average for the subsequent fit.

\begin{table}[]
\caption{Time bins interval for the variability likelihood analysis and corresponding phase, using \cite{2005AJ....129.2018C} ephemeris.}
\centering
\begin{tabular}{c|ccc}
\hline
\hline
\noalign{\smallskip}
Bin & MET Start & MET Stop & $\eta$ Carinae orbital phase \\
\noalign{\smallskip}
\hline
\noalign{\smallskip}
1 & 239557417 & 263317417 & [0.921,1.057]\\
2 & 263317417 & 287077417 & [1.057,1.193]\\
3 & 287077417 & 321608617 & [1.193,1.390]\\
4 & 321608617 & 356139817 & [1.390,1.588]\\
5 & 356139817 & 390671018 & [1.588,1.785]\\
6 & 390671018 & 414431018 & [1.785,1.921]\\
7 & 414431018 & 438191018 & [1.921,2.057]\\
8 & 438191018 & 461951018 & [2.057,2.193]\\
\noalign{\smallskip}
\hline
\end{tabular}
\label{tab:time_bin}
\end{table}

We then refined our analysis by dividing the photon sample into more time intervals. The duration of the bins was chosen to obtain a clear detection (TS $\gtrsim$ 16) of $\eta$ Carinae in all time intervals and to obtain a sequence repeating from an orbit to the next. We split the light curve into two periods, one for periastron and one for apastron, and subsequently we applied a static binning in each of them. The final eight time bins are reported in Table~\ref{tab:time_bin}.

We attributed to each photon a phase calculated from periastron times using ${\rm JD} = 2450799.792+{\rm N_p}\cdot(2024\pm 2)$ \citep{2005AJ....129.2018C}, where ${\rm N_p}$ counts the successive periastrons. A more recent analysis \citep{2008MNRAS.384.1649D} suggests a shorter orbital period of $2022.7 \pm 1.3$ days, but such a small variation does not have any impact on our results.
We then ran another likelihood analysis for each time bin, but given the shorter exposure, we detected a very low significance for the source J1043 in the 4th, 5th, and 8th bin and, consequently, we decided to fix its contribution in those bins to the average value reported in the 3FGL catalogue. 

\begin{figure}[]
\resizebox{\hsize}{!}{\includegraphics{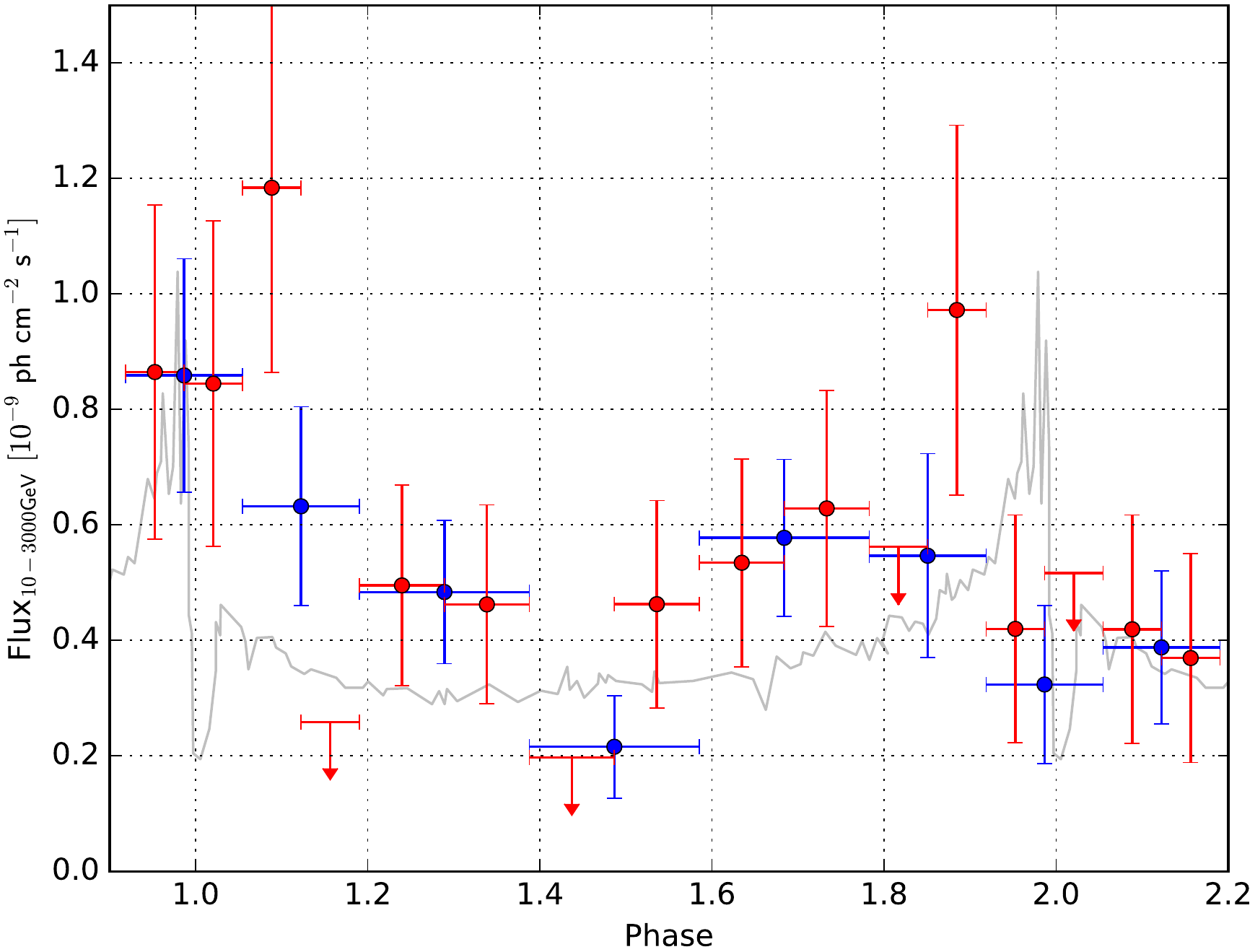}}
\caption{Seven-year high-energy flux light curve of \object{$\eta$ Carinae} obtained from the binned analysis, using the binning as reported in Table \ref{tab:time_bin} (blue points) and with a smaller binning (red points). Error bars are $1\sigma$ and superposed upper limits are 95\%. For comparison we plotted an arbitrarily rescaled X-ray light curve (grey line).}
\label{fig:HE_lc}
\end{figure}

The light curve of the high-energy flux of $\eta$ Carinae obtained from the likelihood analysis is reported in Fig.~\ref{fig:HE_lc}. After the first periastron passage of 2009 the flux of \object{$\eta$ Carinae} decreased slightly towards apastron. The flux did however not increase again toward the periastron of 2014. The last two bins of this light curve can be directly compared with the first two, having the same exposures and orbital phases. In order to search for any faster variability we reduced the temporal size of the bins, which confirmed the absence of any excess during the second periastron.

When performing the fit in each bin, letting free the spectral index $\Gamma$, we can observe that the resulting values are constant within the uncertainties. When performing the same analysis fixing $\Gamma$ to its average value, we found that the resulting fluxes were changing by a few percent at most, i.e. much less than the statistical uncertainties. Therefore the light curve presented in Fig. \ref{fig:HE_lc} does not depend significantly on the exact spectral shape assumed.

Given the relatively low event statistics at high energy, as a countercheck we also performed an unbinned analysis \citep{1996ApJ...461..396M} on the same sample of photons. The results that we obtained are F$_{\rm 10-300GeV}=(4.74\pm0.49) \times 10^{-10}$ ph cm$^{-2}$ s$^{-1}$ and index $\Gamma=-2.29 \pm 0.15$. Both the flux light curve and the spectral index trend are in very good agreement with the results of the binned analysis. For comparison we also ran an analysis with a smaller bin as reported in Fig~\ref{fig:HE_lc}. These results show a bigger error but are consistent with the previous analysis within the uncertainties, even if on a few occasions the smaller statistics did not yield a firm detection of the source and only reached a $95\%$ upper limit.

\begin{figure*}[]
\resizebox{\hsize}{!}{\includegraphics{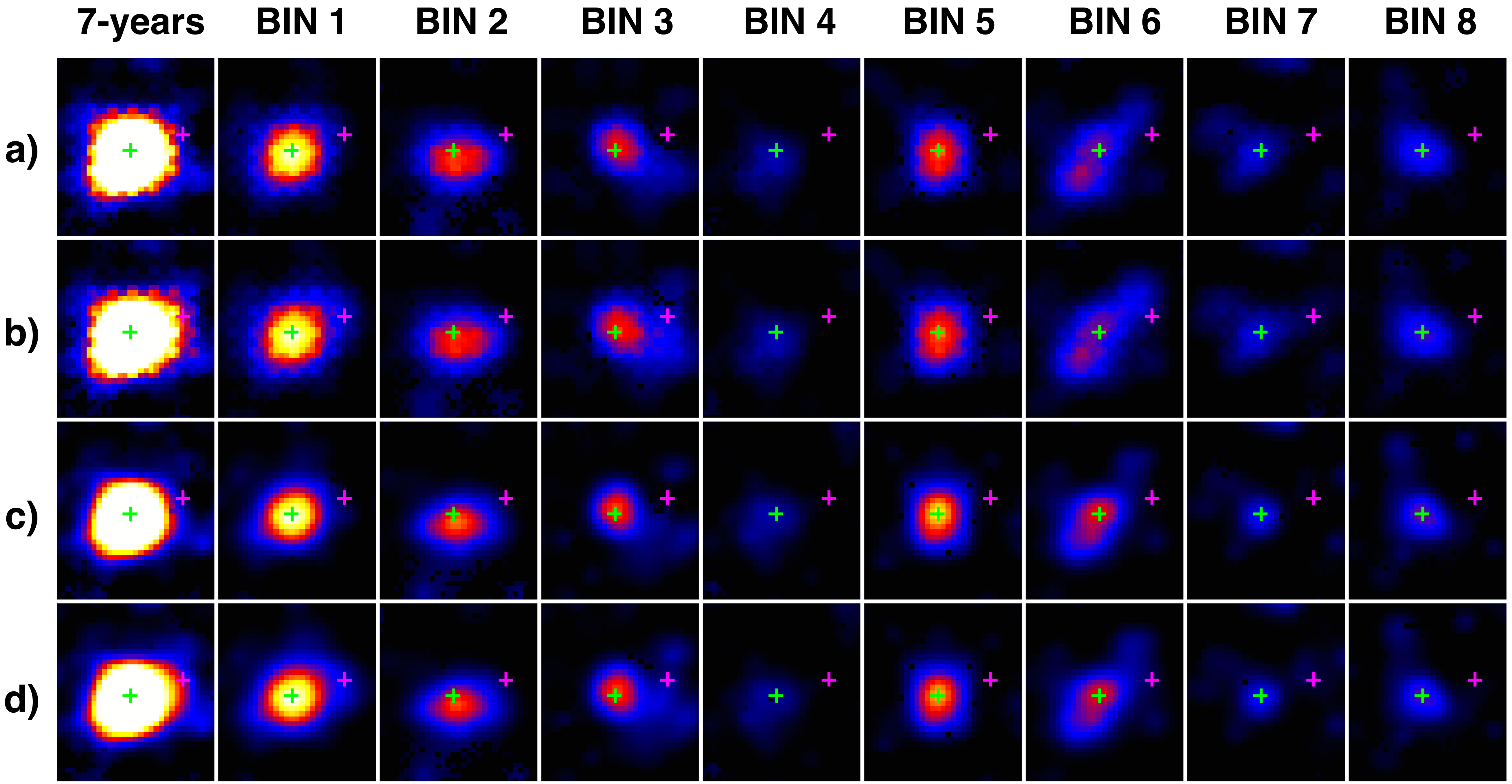}}
\caption{High-energy TS maps for each of the time intervals, corrected for the small differences of exposure times between the 8 phase bins such that the images illustrate the source variability. Each image has the same width $(0.77^{\circ})$. Rows a) and b) show TS maps obtained from the binned analysis, respectively including and excluding J1043 from the model. Rows c) and d) represent the same as a) and b) but from the unbinned analysis. The linear colour map spans TS from 0 to 100. The green and purple crosses are the positions of $\eta$ Carinae and J1043.}
\label{fig:tsmap}
\end{figure*}

We gave special attention in our analysis to J1043. In the Fermi 3FGL catalogue a relatively hard spectrum is reported for that source. At low energy its flux is more than one order of magnitude lower than that of \object{$\eta$ Carinae}, and it reaches nearly one-third of that flux above 10 GeV. Given its small distance ($\lesssim 2r_{68}$ PSF), we analysed the impact that a wrong representation of this source could have on the flux of \object{$\eta$ Carinae}. To set an upper limit on the possible systematic error introduced, we considered the following two cases. We first ran a likelihood analysis keeping the J1043 parameters frozen to its seven-year average value, which gave us a light curve for \object{$\eta$ Carinae} with flux values reduced by $5\%$. Then as a counter check, we completely removed J1043 from the model, obtaining a biased light curve for \object{$\eta$ Carinae} with values up to $11\%$ higher. All those results are in agreement with the TS maps we obtained for each single bin (see Fig.~\ref{fig:tsmap}).

\begin{table}[]
\caption{Best fit coordinates of \object{$\eta$ Carinae} obtained from \texttt{gtfindsrc} with $1\sigma$ uncertainties. Distance is referred to the nominal position of \object{$\eta$ Carinae}.}
\centering
\label{my-label}
\begin{tabular}{c|ccccc}
\hline
\hline
\noalign{\smallskip}
Bin & R.A. & Dec & $1\sigma$ $[']$ & Dist $[']$ & TS \\
\noalign{\smallskip}
\hline
\noalign{\smallskip}
7 years & 161.28 & -59.70 & 0.6 & 0.8 & 368 \\
\noalign{\smallskip}
\hline
\noalign{\smallskip}
1 & 161.28 & -59.71 & 1.3 & 1.5 & 92 \\
2 & 161.23 & -59.72 & 1.5 & 2.4 & 51 \\
3 & 161.29 & -59.67 & 1.3 & 1.1 & 59 \\
4 & 161.39 & -59.71 & 4.0 & 4.0 & 18 \\
5 & 161.29 & -59.68 & 1.2 & 0.9 & 67 \\
6 & 161.30 & -59.69 & 1.9 & 1.0 & 34 \\
7 & 161.27 & -59.71 & 1.8 & 1.4 & 20 \\
8 & 161.30 & -59.69 & 2.2 & 1.2 & 29 \\
\noalign{\smallskip}
\hline
\end{tabular}
\label{tab:gtfindsrc}
\end{table}

\begin{figure}[]
\resizebox{\hsize}{!}{\includegraphics{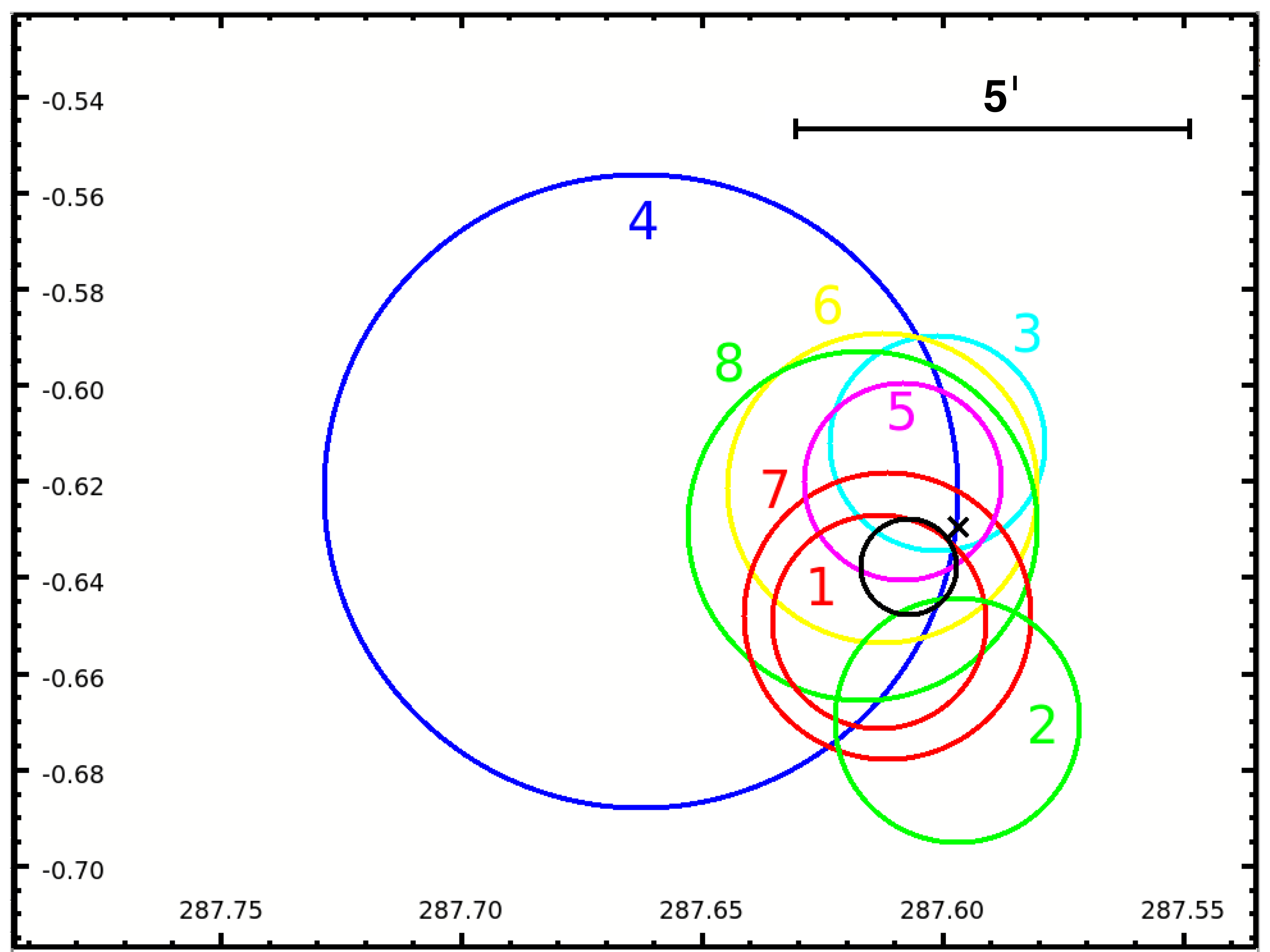}}
\caption{Error circles of $1\sigma$ for $\eta$ Carinae derived from the analysis of each time bin. Labels identify the time bin as in Table \ref{tab:time_bin}; black circle refers to the whole 7 years of data. The black cross shows the nominal position of $\eta$ Carinae. As for Fig.~\ref{fig:tsmap}, the image is in Galactic coordinates with North up and longitude increasing towards the left.}
\label{fig:gtfindsrc}
\end{figure}

The average seven-year TS maps perfectly match the PSF of the instrument, while in the shorter time bins the emission from \object{$\eta$ Carinae} is often broadened. So far in our analysis we have always kept fixed the location of \object{$\eta$ Carinae}, using its nominal coordinates. Exploiting the unbinned analysis and the \texttt{gtfindsrc} tool, we left the spatial coordinates as a free parameter in the fit and looked for the best coordinates to maximize the likelihood. Such an analysis has been performed in each single bin. The results are shown in Fig.~\ref{fig:gtfindsrc} and reported in Table~\ref{tab:gtfindsrc}. The eight error circles of \object{$\eta$ Carinae} are represented for each time bin, while the black circle indicates the result obtained running \texttt{gtfindsrc} on the whole seven-year data sample. The radius of each circle represents a $1\sigma$ error. The nominal position of \object{$\eta$ Carinae}, shown by the black cross, is very well in agreement with the results of the Fermi analysis, being within the error circle more than 78$\%$ of the time.

\subsection{Low-energy emission: 0.3 - 10~GeV} \label{sec:low_energy}

Extending the analysis to lower energies, the PSF becomes broader and the effective area, acceptance, and energy resolution worsen, making the analysis more challenging. In particular it requires us to enlarge the ROI and consequently to increase drastically the number of sources, the number of free parameters, the uncertainties (also related to the non-perfect Galactic diffuse emission model), and the computation requirements. For these reasons we chose a lower bound for our analysis of 300 MeV, as a compromise. This choice also keeps the flux systematic uncertainty below $5\%$\footnote{http://fermi.gsfc.nasa.gov/ssc/data/analysis/documentation/\\/Pass8\_edisp\_usage.html}.

As the photon statistic is much better at lower energies, we used only a binned analysis, similar to that described in Sect. \ref{sec:high_energy} but with the following adaptations. As the total (front+back) $95\%$ containment angle above 300 MeV is now smaller than $7^{\circ}$ (on-axis), we took into account all photons arriving within $14^{\circ}$ from the nominal position of $\eta$ Carinae and created a sky model including all sources in the ROI enlarged by $7^{\circ}$. We let the normalization parameters free to vary for all sources within $7^{\circ}$ from $\eta$ Carinae with an average TS $>$100 or presenting a variability index higher than 72.44 (see Sect.~\ref{sec:high_energy} for explanation). As the flux of J1043 is much smaller than that of $\eta$ Carinae at low energy, we kept frozen all its parameters to the values given in the 3FGL catalogue. Even though the extended pulsar wind nebula HESS J1303-631 lies more than $16^{\circ}$ away from $\eta$ Carinae, we included an appropriate extended\footnote{http://fermi.gsfc.nasa.gov/ssc/data/access/lat/4yr$\_$catalog/\\/LAT$\_$extended$\_$sources$\_$v15.tgz} representation in the model. We checked again in the FAVA \citep{2013ApJ...771...57A} weekly flare list and found seven events within our ROI and verified that all were effectively associated with sources whose normalization were left free to vary.

To obtain a rough idea of the spectral shape of $\eta$ Carinae, we started to split our analysis into separate energy bins, where the spectrum could be approximated locally as a simple PL. At the same time we kept a sufficient statistic in order to detect our source with at least $5\sigma$. We defined five logarithmically equal energy bins, from 300 MeV up to 100 GeV, and performed a separate likelihood analysis in each of them for the different orbital phase intervals. In the most energetic bin [30~GeV-100~GeV] $\eta$ Carinae did not always reach the required TS, and in these cases we merged the 4th and 5th energy bins together in the analysis. We computed the integral energy flux for each bin, converted them to luminosities (assuming a distance of 2.3 kpc), and plotted them in Fig.~\ref{fig:espec} (later described in Sect.~\ref{sec:simulation}) for two orbital phase bins (0.92-1.05 and 0.39-0.59), corresponding to periastron and apastron, respectively. In these plots, the centre of each point is not a logarithmic average, but is computed making a weighted average using the energy dependent function resulting from each fit as a weight. The result of this analysis indicates that the low-energy spectrum of $\eta$ Carinae features some curvature that could be represented locally, for example by a cutoff power law or a broken power law and that an excess could be observed in some spectra above 10 GeV.

We therefore performed the complete analysis, in band 0.3-10 GeV, assuming a power law spectrum with an exponential cutoff (PLEC) dN/dE = N$_{\rm 1GeV} {\rm (E/1~GeV)}^\Gamma {\rm e}^{\rm -E/E_c}$ for $\eta$ Carinae, where ${\rm N}_{\rm 1GeV}$ is the normalization in units of ph~cm$^{-2}$~s$^{-1}$~MeV$^{-1}$, $\Gamma$ is the power law index, and E$_c$ is the cutoff energy.

To perform the analysis in the orbital phase bins we always fixed the Galactic and extragalactic diffuse emission normalization to their seven-year averaged values (N$_{\rm Gal}=0.962 \pm 0.002$ and N$_{\rm Exgal} = 1.20 \pm 0.03$). Leaving these two normalizations free affects the flux of $\eta$ Carinae by less than $5\%$. The difference of normalization obtained between the high and low-energy analysis for the diffuse emission is related to the very different sizes of the respective ROI.

\begin{figure}[b]
\resizebox{\hsize}{!}{\includegraphics{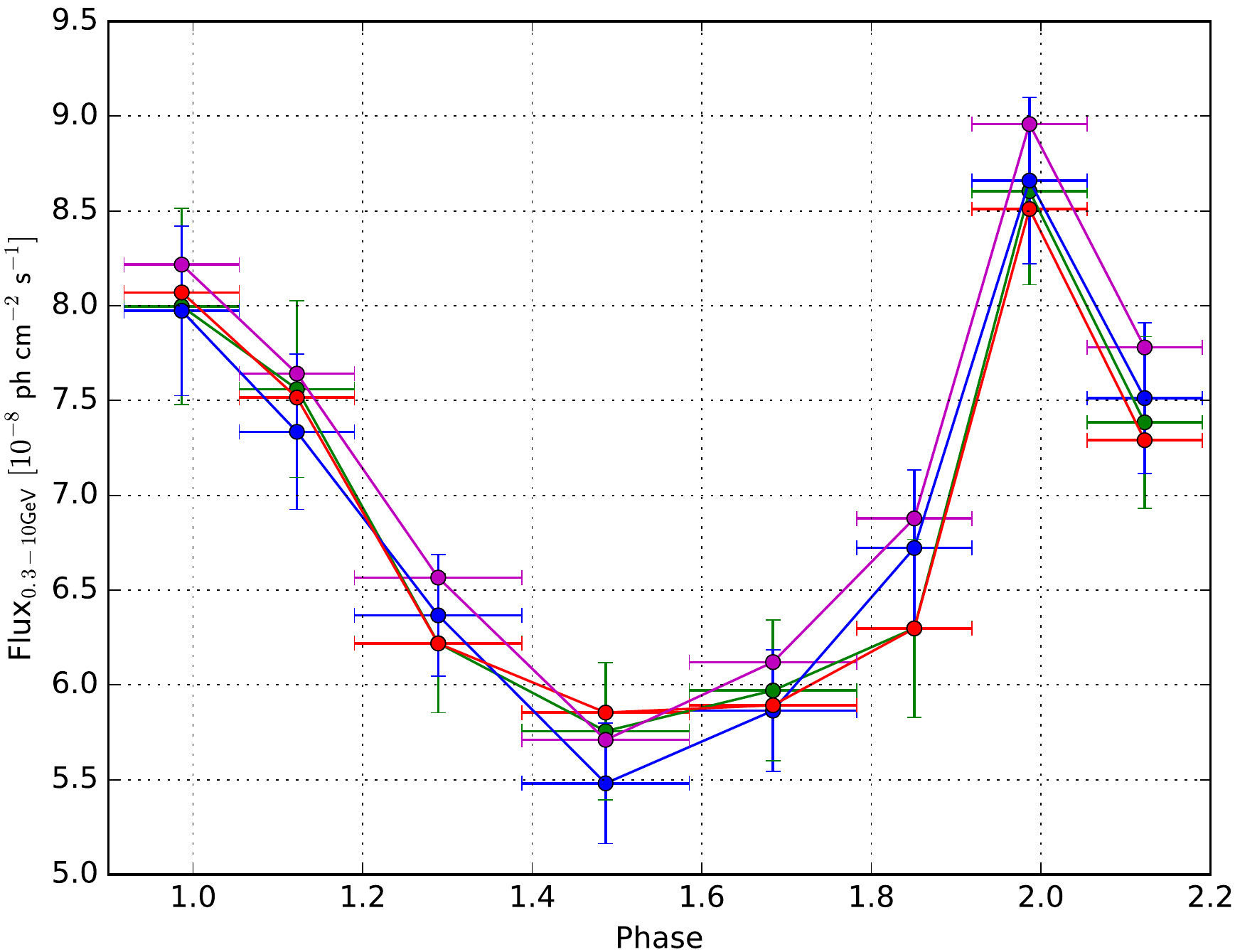}}
\caption{Integrated flux light curve of $\eta$ Carinae assuming different spectral models: PLEC with three (green) or two (blue) free parameters and BPL with three (red) or two (magenta) free parameters. Error bars are $1\sigma$.}
\label{fig:PLEC}
\end{figure}

The values of the photon index and energy cutoff are locally correlated and not very meaningful. Figure~\ref{fig:PLEC} compares the integrated flux obtained with the PLEC model with all parameters free or fixed to the average of $\Gamma_{PLEC}=-2.12$ and with a broken power law (BPL) model with the low-energy spectral index left free or fixed to the average of $\Gamma_{BPL}=-2.14$. The low-energy $\gamma$-ray flux light curve does not change significantly and is therefore a good measure of the emission of $\eta$ Carinae. Figure~\ref{fig:simul} shows the combined result of the binned high-energy analysis obtained in Sect.~\ref{sec:high_energy} together with the results derived here for the low-energy band (assuming a distance of 2.3 kpc).

As the data above 10 GeV of the second periastron fails to reproduce the high flux levels detected during the first periastron, the light curve of the low-energy band is now of prime importance to confirm the orbital variability of the $\gamma$-ray emission. The probability of obtaining an orbital flux variation by chance, as reported in Fig.~\ref{fig:PLEC}, is lower than 5$\times$10$^{-9}$ (5.9$\sigma$).

Finally, as the low-energy $\gamma$-ray variability appears similar for the two periastrons (see Fig.~\ref{fig:PLEC}) and the spectra are well compatible within $2\sigma$ (see Fig.~\ref{fig:espec}), we used the good statistics available to perform a merged analysis of these two periods on even shorter time bins. We merged the data covering the period from phase [0.92-1.19] and [1.92-2.19] and split the data according to phase bin intervals of 2\% (i.e. $\sim$ 40 days). We performed a single likelihood analysis in each time bin, representing the emission of $\eta$ Carinae with a simple PL. $\eta$ Carinae was detected above 12$\sigma$ in every bin, increasing the detection significance up to 50\% compared to a single periastron analysis. We also performed the analysis by shifting the central phase point of each bin by half of its width to obtain a better sampling of the variability. The results of both analyses are shown in Fig.~\ref{fig:peri}. We also performed another similar analysis increasing the lower energy threshold to 600 MeV and 1GeV to reduce the energy range, trying to exploit the better PSF at higher energy and increasing at the same time the robustness of the model approximations. The results showed exactly the same variability trend and similar likelihoods were obtained. We added in Fig.~\ref{fig:peri} the results obtained previously for two broad bins adjacent to the periastron period.

\section{Comparison with simulations} \label{sec:simulation}

\cite{2011ApJ...726..105P} presented three-dimensional hydrodynamical simulations of $\eta$ Carinae including radiative driving of the stellar winds \citep{1975ApJ...195..157C}, optically thin radiative cooling \citep{2000adnx.conf..161K}, gravity, and orbital motion. The main aim of these simulations was to reproduce the X-ray emission by analysing the emissivity and self-obscuration of the stellar wind. The simulations reproduced the observed X-ray spectra and light curves reasonably well, excepting the post-periastron extended X-ray minimum, where flux was overestimated and the wind collision disruption was inhibited. Additional gas cooling, for example by particle acceleration and inverse-Compton processes, could increase the cooling and disruption of the central wind collision zone.

\cite{2011ApJ...726..105P} provided us with the results of their simulations, i.e. temperature, density, and three-dimensional velocities in the cells of the adaptive mesh for various orbital phases. To estimate the non-thermal emission we first calculated the maximum energies that could be reached by electrons and hadrons \citep[as in][]{2011A&A...526A..57F} cell by cell assuming a di-polar magnetic field at the surface of the main star, perpendicular to the orbital plane (reality is probably more complex with the two stars contributing). The magnetic field is the only additional parameter, which can be tuned. We calculated shock velocities and mechanical power in every cell, including those outside the shock region. As expected, most of the shock power is released on both sides of the wind collision zone and in the cells downstream the wind-collision region \citep{2006ApJ...644.1118R}. The increasing shock area compensates for the loss of the released energy density up to a relatively large distance from the centre of mass, explaining why the X-ray luminosity at apastron is about a third of the peak emission at periastron. 

The energy available in electrons and hadrons were then summed in the ranges 0.3 $< \rm{E}_e <$ 10 GeV and $\rm{E}_p>20$ GeV, respectively, to match the spectral bands observed by Fermi-LAT. The local cell physical properties can be used to easily estimate pion production as long as the Larmor radius is similar to the cell size. The minimum size of the cells in the simulation is $\sim10^{11}$ cm, which is larger than the proton Larmor radius for Lorentz gamma factor up to 10$^5$. Only one-third of the power accelerating protons is available to produce $\gamma$-rays through the neutral pion channel. Electron cooling and pion decay occur instantaneously when compared to other timescales.

To consider the possible effects of photon-photon opacity we calculated the X-ray thermal emission in each cell and evaluated the optical depth along different lines of sight. As the current orientation of the binary system, with respect to the Earth, still presents some uncertainties \citep{2012ApJ...746L..18M}, we used several possible directions; this provided optical depth $\tau$ that varied between $\sim10^{-6}$ at apastron and $\sim10^{-2}$ at periastron. This excludes explaining the 1-100 GeV spectral shape by the effects of photon-photon absorption \citep{2012A&A...544A..98R}.

Thermal emission increases towards periastron. The mechanical luminosity available in the shock also increases towards periastron and almost doubles in the phase range $\approx1.05-1.15$. The latter peak corresponds to a bubble with reverse wind conditions developing because of the orbital motion, effectively doubling the shock front area during about a tenth of the orbit \citep[see Fig. 9 of][]{2011ApJ...726..105P}. The density of this bubble is low so its thermal emission $(\propto \rm{density}^2)$ does not contribute significantly to the X-ray light curve. The mechanical luminosity shows a local minimum between phases 1.0 and 1.05 when the central part of the wind collision zone is disrupted. 

\begin{figure}[]
\resizebox{\hsize}{!}{\includegraphics{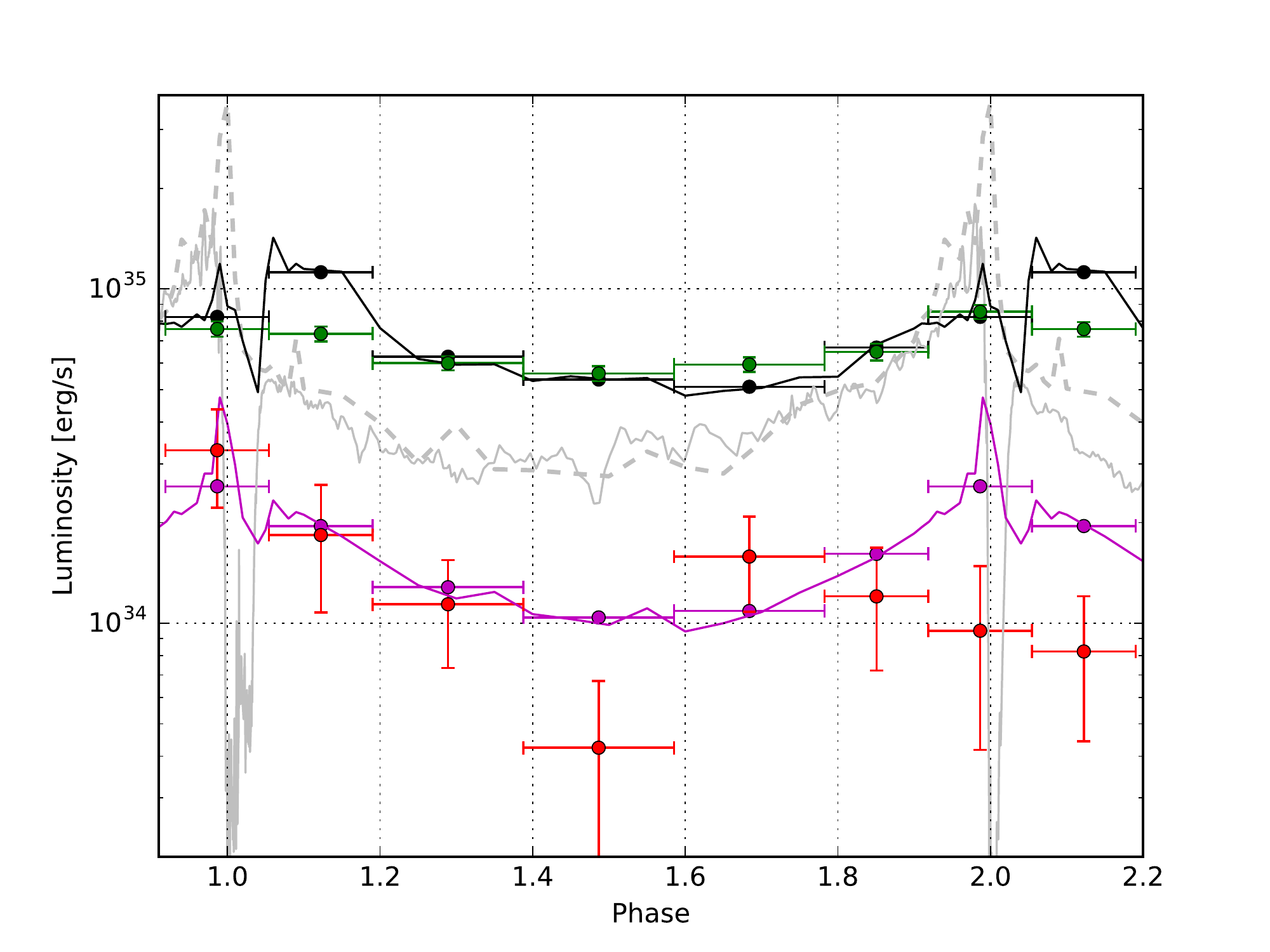}}
\caption{Simulated and observed X-ray and $\gamma$-ray light curves of $\eta$ Carinae. The black and purple lines and bins show the predicted inverse-Compton and neutral pion decay light curves. The green and red points show the observed Fermi-LAT light curves at low (0.3-10 GeV) and high (10-300 GeV) energies. The dim grey light curves show the observed (continuous) and predicted (dash, without obscuration) thermal X-ray light curves. Error bars are $1\sigma$.}
\label{fig:simul}
\end{figure}

Electron cooling, through inverse-Compton scattering, is very efficient and such $\gamma$-rays are expected to peak just before periastron. A secondary inverse-Compton peak could be expected above phase 1.05, although its spectral shape could be very different as the UV seed thermal photons are of lower density when compared to the location of the primary shock close to the centre of the system. In our simplified model we assumed that the spectral shape of the seed photons is the same in all cells of the simulation (r$^{-2}$ dependency is taken into account), and that these soft photons are sufficient to cool down all the relativistic electrons. The relative importance of the second peak, however, depends on the magnetic field geometry; radiation transfer, which is neglected in our model; obscuration; and details of the hydrodynamics, which do not represent the soft X-ray observations very well in this phase range. These details are not well constrained by the available observations and we did not try to refine them. 

The situation is different for hadrons. Unless the magnetic field is very strong ($>$ kG) hadronic interactions mostly take place close to the centre and a single peak of neutral pion decay is expected before periastron.

Figure \ref{fig:simul} shows the X and $\gamma$-ray light curves predicted by the simulations for a magnetic field of 500 G and assuming that 1.5\% and 2.4\% of the mechanical energy is used to accelerate electrons and protons, respectively. To ease the comparison between observations and simulations, the results of the latter were binned in the same way as the observed data.

The thermal X-ray emission matches the observations pretty well \citep[by construction,][]{2011ApJ...726..105P}. For the simulated curve in Fig. \ref{fig:simul}, we use the orbit-IA model by Parkin, which uses instantaneous acceleration and not radiative driving of the winds. In addition, the thermal X-ray light curve does not take self-obscuration into account and therefore does not match the observations around periastron. The predicted $\gamma$-ray emission induced by the hadrons and electrons are also at the right level, although significant differences exist between simulations and observations.

\begin{figure}[b]
\resizebox{\hsize}{!}{\includegraphics{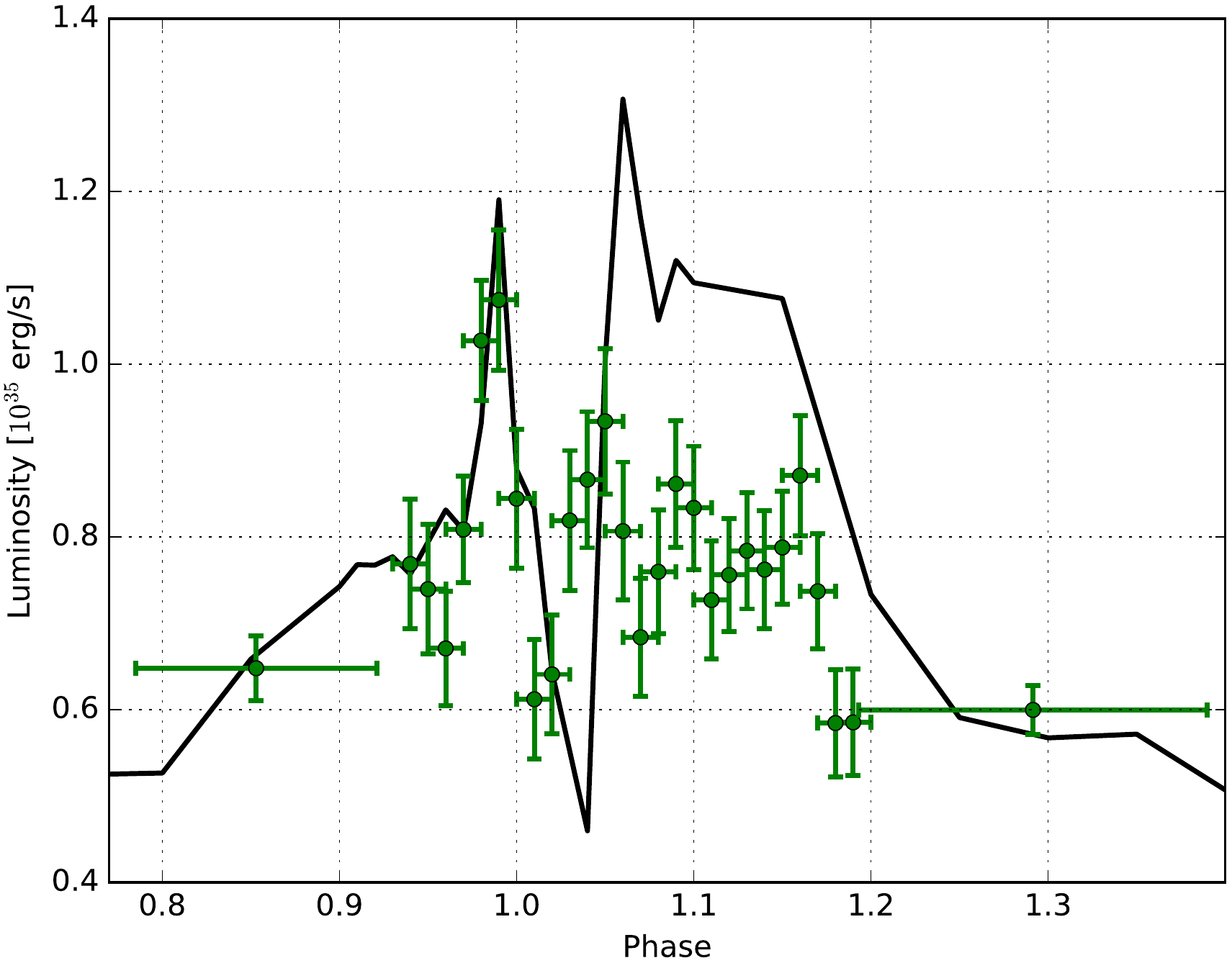}}
\caption{Merged Fermi LAT analysis (0.3-10 GeV) of the two periastrons for narrow time bins. The two broad bins and the black curve are the same as in Fig. \ref{fig:simul}.}
\label{fig:peri}
\end{figure}

\begin{figure*}[ht]
\resizebox{\hsize}{!}{\includegraphics{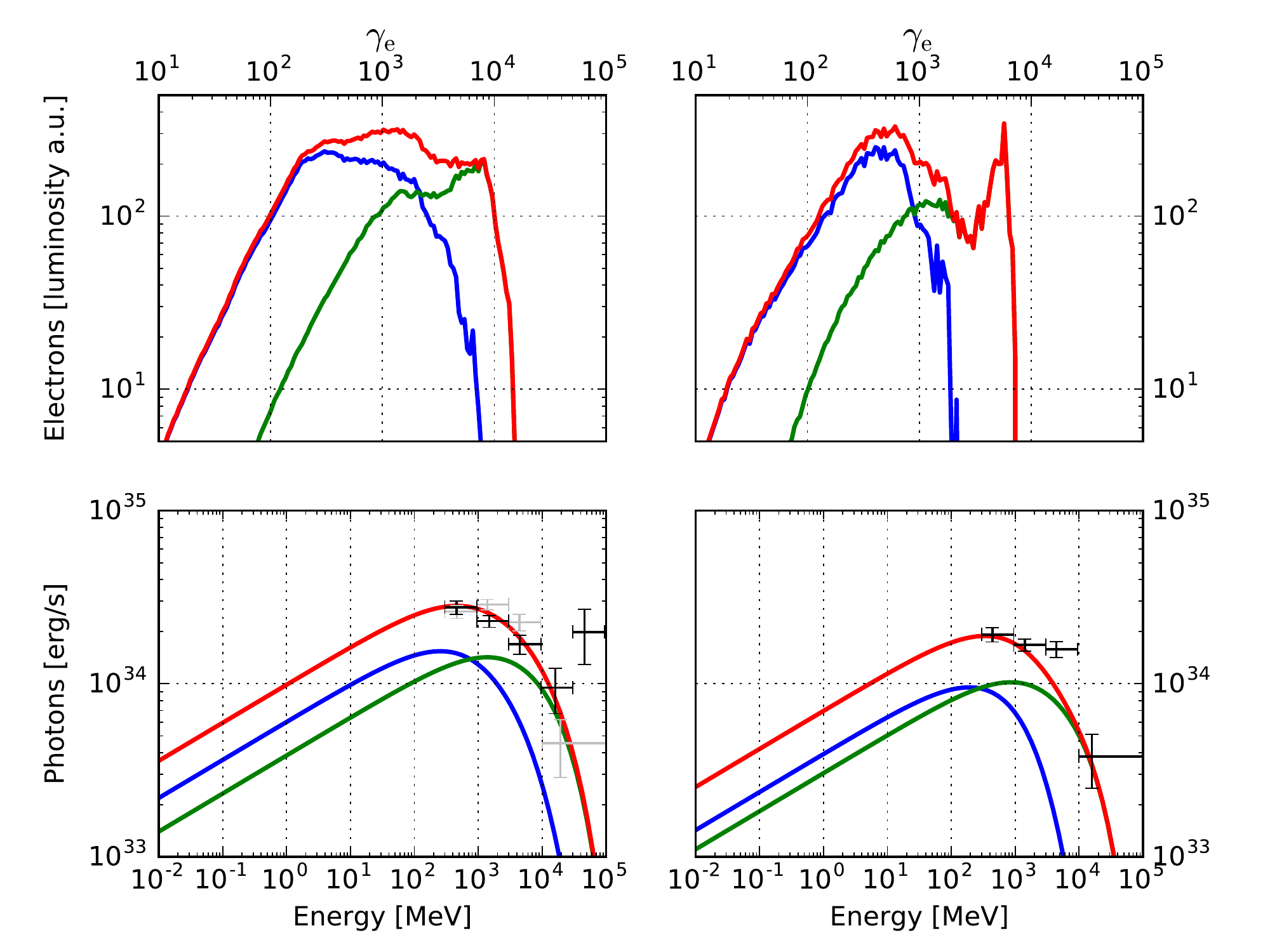}}
\caption{Electrons and photons luminosity spectra at periastron (phase 0.92-1.06; left) and at apastron (phase 0.39-0.59; right). The top panels show the spectra (arbitrary units) of the electrons accelerated in the wind of the primary (green) and secondary (blue) stars and their sum (red). The lower panels show the inverse-Compton emission of both components and the total emission, under the highly simplified assumption that the inverse-Compton parameters (geometry and soft photon spectra) are the same in all cells. The black and grey points are the broadband fluxes derived from Fermi data for the first and the second orbital cycle, respectively. The simulation results were averaged over the orbital phase range corresponding to the periastron observation, as the electron spectra vary quickly during that interval.}
\label{fig:espec}
\end{figure*}

Both the predicted inverse-Compton emission and the observed (0.3-10 GeV) LAT light curve show a broad peak extending on both sides of periastron, as expected from the evolving shock geometry. The amplitude of the variability in the simulation depends on the number/size of those cells where particles can be accelerated up to relevant energies, which in turn depends on the magnetic field. Probing the range suggested by \cite{2012SSRv..166..145W}, a surface magnetic field that is larger than 400~G provides a good match to the observations, while lower fields produce variations that are too large. In this work we did not considered any magnetic field amplification at the shock, which in turn could obviously scale down the surface magnetic field required to get equivalent results. Assuming a field of 500 G for the rest of the discussion, the predicted flux at phase 1.1 is two times larger than observed. This discrepancy largely comes from the energy released in the inverted wind bubble after periastron. The ratio of the emission generated in the shocks on both sides of the wind collision zone is relatively constant along the orbit except at phase 1.1, where much more power is generated in the shock occurring in the wind of the secondary star. The inverted bubble might either be unstable in reality or might produce a significantly different inverse-Compton spectrum.

Relativistic electrons immersed in such a high magnetic field produce a synchrotron radiation at low energy. The ratio of the energy that electrons lose via synchrotron and inverse-Compton processes is equal to the ratio of the magnetic field energy density over the photon field energy density, i.e. $P_{synch}/P_{IC}=U_B/U_{rad} = B^2 4\pi R^2 c/8\pi L \approx 7.8 \times 10^{-31} \cdot B[G]^2 \cdot R[cm]^2$. If we know the inverse-Compton spectrum, we can estimate the synchrotron peak luminosity, which around apastron results to be several orders of magnitude ($\sim10^6$) fainter than the inverse-Compton peak. The synchrotron emission peak should reach its maximum in the optical band only very close to periastron, as it is only two orders of magnitude fainter than the inverse-Compton peak. Those limits are in agreement with the estimated radio upper limit \citep{2003MNRAS.338..425D}.

Since the low-energy spectra during both periastrons are sufficiently in agreement (see Fig.~\ref{fig:espec} described later), we analysed simultaneously the Fermi LAT low-energy data derived from the two periastrons, binned in shorter time intervals (Fig.~\ref{fig:peri}). These data show a peak at periastron, a minimum at phase 1.02, and a second broad peak at phase 1.1. This is very similar to the prediction of the simulation for the inverse-Compton luminosity. The only notable exception is that the observed second broad peak is slightly shifted towards earlier phases and has a lower luminosity when compared to the simulation. The similarities between the observations and the simulation for the $\gamma$-ray peak and minimum with consistent duration and amplitude are very encouraging. The phase difference could be related to the eccentricity $(\epsilon=0.9)$ assumed in the simulation, which is not well constrained observationally \citep{2000ApJ...528L.101D,2001ApJ...547.1034C}, and this has an important effect on the inner shock geometry. 

Figure \ref{fig:espec} shows that the distribution of $\gamma_e$, weighted by the emissivity, is relatively smooth and that the expected photon distribution is very smooth. The difference in the electron spectral shape on both sides of the wind collision zone cannot explain the two components $\gamma$-ray emission as suggested by \cite{2011A&A...530A..49B}, who assumed a simplified geometry. We obtain a good match between the observed low-energy $\gamma$-ray spectrum and the predictions of the simulations at periastron, even though some discrepancy can be observed at apastron where an excess is observed between 2 and 10 GeV.

The inverse-Compton emission peaks slightly below 1 GeV and does not extend beyond 10 GeV at a level that is consistent with the observations during the first periastron in contrast with the conclusions from \cite{2015MNRAS.449L.132O}, which attribute the full Fermi LAT detection to hadronic emission. Their simulations predict a smaller variation between periastron and apastron, a longer flare around periastron, and a deeper minimum when compared to the observed data. Such discrepancies might be due to the simplified geometry assumed by the authors and by the artificially reduced particle acceleration at periastron. Inverse-Compton emission and neutral pion decay \citep{2011A&A...526A..57F} remains therefore a very good candidate to explain the Fermi observations. The fraction of the shock mechanical luminosity accelerating electrons appears to be slightly smaller than the fraction that accelerates protons. These results differ from the efficiencies derived from simulations of particle acceleration in supernova remnants \citep{2015PhRvL.114h5003P}, but those simulations involve low magnetic field, radiation energy, and particle densities, i.e. very different physical conditions than found in $\eta$ Carinae. An instrument sensitive in the 1-100 MeV band would be able to discriminate between our model and the one proposed by \cite{2015MNRAS.449L.132O}.

The simulated pion induced $\gamma$-ray light curve and its variability amplitude show a single peak of emission centred at periastron, which is in good agreement with the Fermi LAT observations of the first periastron. The results of the observations of the second periastron are different in that they have a weaker emission. It has been suggested that the change of the X-ray emission after that periastron, where a significant decrease can be observed in Fig. \ref{fig:simul}, \citep[see also ][]{2015arXiv150707961C}, was the signature of a change of the wind geometry possibly because of cooling instabilities. A stronger disruption or clumpier wind after the second periastron could perhaps induce a decrease of the average wind density and explain that fewer hadronic interactions and fewer thermal emission took place without affecting inverse-Compton emission much.

\begin{figure}[h]
\resizebox{\hsize}{!}{\includegraphics{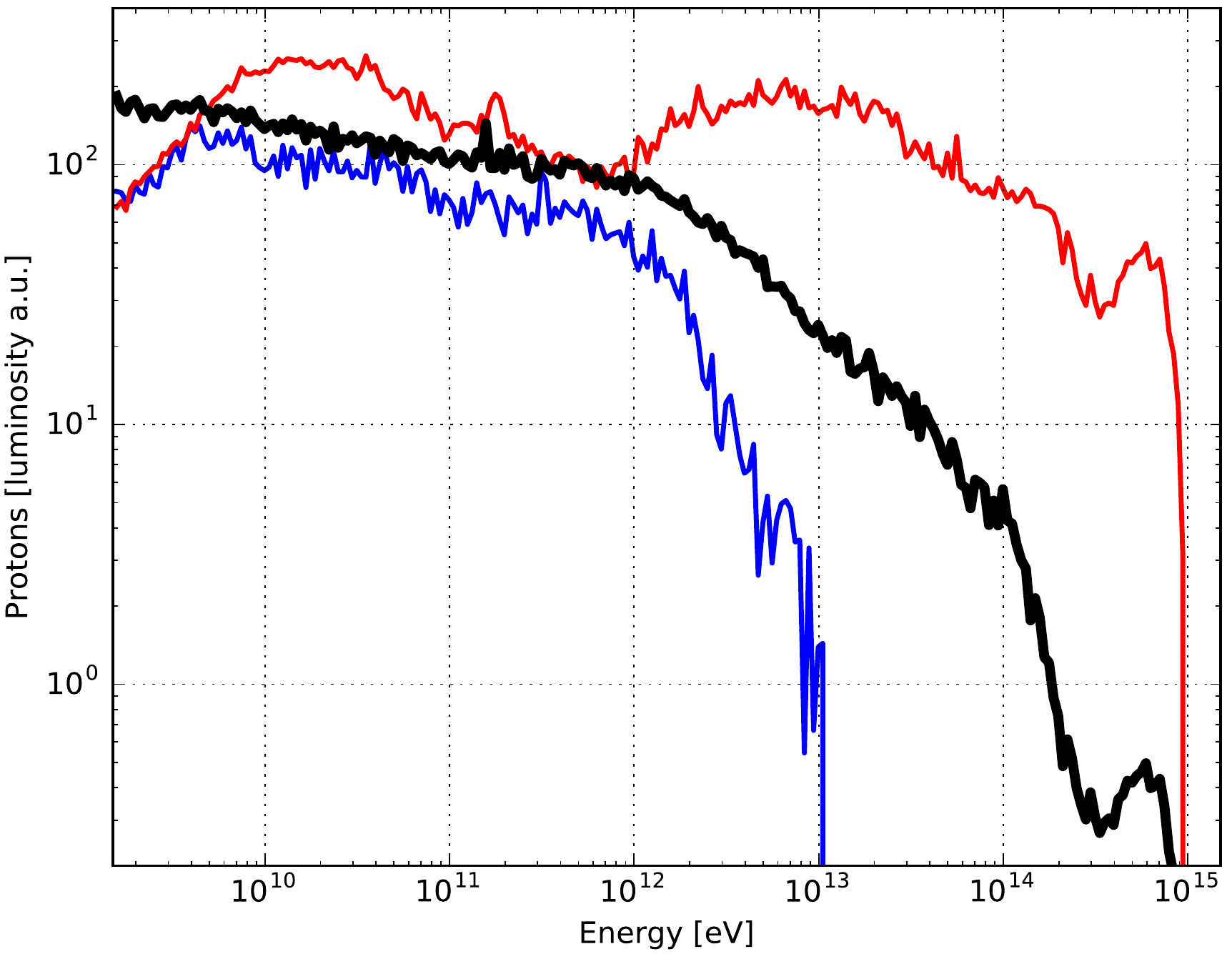}}
\caption{Protons luminosity spectra (arbitrary units) at periastron (red; phase 1.0), apastron (blue; phase 0.5) and accelerated on average along the orbit (black). }
\label{fig:protons}
\end{figure}

Figure~\ref{fig:protons} shows the proton spectra obtained from the simulation at apastron, periastron, and averaged over the orbit. Protons could be accelerated up to $10^{15}$ eV around periastron and reach $10^{14}$ eV on average. 
The choice of a lower magnetic field reduces those energies at apastron to $\sim6\times 10^{12}$~eV and $\sim 2 \times 10^{12}$~eV and at periastron to $\sim 5.6 \times 10^{14}$~eV and $\sim 1.9 \times 10^{14}$~eV for 300~G and 100~G, respectively. $\eta$ Carinae can therefore probably accelerate particles close to the knee of the cosmic-ray spectrum. The spectra and the maximum particle energy depend of course on several assumptions, in particular the magnetic field. The highest energy $\gamma$-rays are photo-absorbed and orbital modulation could be expected in the TeV domain. The duration of the periastron bin [0.92-1.06] corresponds to more than 260 days and is longer than the interaction timescale of the protons responsible for the flux variability.

$\gamma$-ray observations can probe the magnetic field and shock acceleration in detail, however the quality of the current data above 1 GeV does not yet provide enough information to test hydrodynamical models including detailed radiation transfer (inverse-Compton, pion emission, and photo-absorption). The interplay between disruption and obscuration does not yet account for the X-ray minimum and orbit to orbit variability. More sensitive $\gamma$-ray observations will provide a wealth of information and allow us to test the conditions and physics of the shocks at a high level of details, making of $\eta$ Carinae a perfect laboratory to study particle acceleration in wind collisions.

\section{Conclusions} 

We have used the hydrodynamic simulation of \cite{2011ApJ...726..105P}, which was developed to reproduce the thermal soft X-ray light curve of $\eta$ Carinae, and have estimated leptonic and hadronic Fermi acceleration, inverse-Compton emission, and neutral pion decay cell by cell assuming a di-polar magnetic field at the surface of the primary star. The results of the simulation were  compared with the light curves and spectra observed by Fermi LAT between mid-2008 and mid-2015. We increased the data sample by $\sim25$\% with respect to previous analyses and exploited the much better performance of the new PASS8 Fermi-LAT pipeline and of the updated instrument responses.
We performed a low-energy and high-energy analysis, from 300 MeV to 10 GeV and from 10 GeV up to 300 GeV, respectively, using the binned \citep{1979ApJ...228..939C} and unbinned analyses \citep{1996ApJ...461..396M}. We used different time bins and also performed a low-energy merged analysis combining data with the same orbital phases, when possible, to increase the signal-to-noise ratio. We looked for high and low-energy flux variability of $\eta$ Carinae and analysed its spectral variations at different orbital phases. 

We found a good match between the accuracy of the simulation, even if simplified, and the signal-to-noise of the observations. The comparison between simulation and observations led to several results.

\begin{enumerate}
\item The centroid of the $\gamma$-ray source observed by Fermi LAT is compatible with the position of $\eta$ Carinae within less than 1 arcmin. The low-energy (0.3-10 GeV) $\gamma$-ray light curve is modulated along the orbit and shows a very similar and highly significant modulation (5.9$\sigma$) during the periastrons of 2009 and 2014, indicating that it is driven by the orbital motion of the system.

\item Around periastron the low-energy (0.3-10 GeV) $\gamma$-ray flux varied by nearly a factor 2 in less than 40 days. A significant fraction of the $\gamma$-rays are therefore emitted by a source smaller than the Homunculus Nebula in contrast with the hypothesis of \cite{2010ApJ...718L.161O}.

\item The maximum over the minimum flux ratio observed at low energy (0.3-10 GeV) is 1.53 considering broad phase bin and 1.92 considering bins of 40 days. This matches the results of the simulations assuming that the magnetic field at the surface of the primary star is larger than $\sim 400$ G. Smaller values of the magnetic field shorten the volume where electrons could be accelerated to sufficient energies, increase the expected variability amplitude beyond the observed one, and decrease the expected $\gamma$-ray luminosity.

\item A surface magnetic field larger than $\sim1$ kG would produce a secondary peak of emission after periastron that is stronger than the periastron peak, which is not observed. A large part of the secondary peak, observed in the data, is linked with a bubble with reversed wind conditions created after periastron and lasting for about a tenth of the orbit. We note that $\gamma$-ray observations together with improved simulations should allow us to constrain the magnetic field in the system even more accurately.

\item The primary maximum observed just before periastron perfectly matches the prediction of the simulation (amplitude, phase, and duration). The secondary peak occurs slightly earlier and with a lower amplitude than predicted. We assume that these discrepancies come from an inaccurate eccentricity and from the extremely simplified treatment of inverse-Compton scattering. The $\gamma$-ray observations should allow us to constrain the eccentricity of the orbit of $\eta$ Carinae more accurately than possible with current optical observations.

\item The amplitude and pattern of the low-energy (0.3-10 GeV) $\gamma$-ray variability correspond in general very well with the predictions. The luminosity of the pion decay depends on the density and a larger variability is expected. The low-energy $\gamma$-rays are therefore very likely to be emitted by inverse-Compton emission in contrast with the claims of \cite{2015MNRAS.449L.132O}.

\item The match between the electron distribution predicted by the simulation and the observed cutoff energy, as well as the negligible photon-photon opacity due to the hot shocked gas in the wind collision as computed along different lines of sight, are strong arguments against the scenario suggested by \cite{2012A&A...544A..98R}.

\item The $\gamma$-ray spectrum observed at apastron shows a discrepancy with the predictions assuming a simplified inverse-Compton treatment. This is very likely indicating that the seed soft photon spectrum is not identical everywhere, as currently assumed by the simulations. Spectral variability therefore provides additional constrains on the shock geometry that can be used by more accurate simulations.

\item The high-energy ($>10$ GeV) $\gamma$-ray component is poorly constrained by the observations. It was well detected during the periastron of 2009, but only weakly detected during the periastron of 2014 and at apastron. The amplitude of variability and the level of the emission however match the expectations for pion decay, inverse-Compton emission is ruled out at such energies, in contrast with the claims of \cite{2011A&A...530A..49B}. Both the high-energy $\gamma$-rays and the thermal X-ray emission were weaker during the second periastron, while the inverse-Compton emission was not affected much. This indicates that something peculiar happened in the densest region of the wind collision zone in 2014. Observation of the next periastrons with Fermi-LAT and by the Cherenkov Telescope Array  (CTA) are required to probe the high-energy component and the wind and shock geometry further through $\gamma$-ray pair conversion.

\item With the constraints derived on the magnetic field, the simulations predict that $\eta$ Carinae should be a Pevatron, as this object is able to accelerate protons nearly up to $\sim10^{15}$ eV. Assuming that for each photon originated via hadronic processes we also have the production of one neutrino, we derive a neutrino flux above 10 TeV that might reach $10^{-9}$ GeV s$^{-1}$cm$^{-2}$ on average, which is of the order of the IceCube neutrino sensitivity for several years of observations \citep{2017ApJ...835..151A}. Stacking some months of periastron data over many orbits should in principle allow the detection of one PeV neutrino, well above the atmospheric background. 

\item The lack of statistics at high energy does not allow us to constrain any physical information about the hadronic spectrum. But it is evident that a pure leptonic scenario is not able to reproduce the high-energy spectrum observed during the first periastron. A strong $\gamma$-ray variability is expected above 100 GeV. Depending on the assumed soft energy photon distribution and the consequent $\gamma$-$\gamma$ absorption at very high energy, $\eta$ Carinae could be detected by the CTA southern array (including four large size telescopes) at more than $10\sigma$ in spectral bins of $\Delta {\rm E/E} = 20\%$ for exposures of 50 hours, which would be sufficient to measure separately the variability along the orbit of the high-energy component and of photo absorption \citep{Acharya20133}. $\eta$ Carinae could yield to $10^{48-49}$ erg of cosmic-ray acceleration which is a number close to the expectation for an average supernova remnant \citep{2016APh....81....1B}. 
\end{enumerate}

$\eta$ Carinae is a wonderful laboratory to study particle acceleration in wind collisions. We have demonstrated that the data from Fermi match the simulation expectations, confirming that Fermi acceleration takes place and providing a new tool to diagnose magnetic fields, shock processes, and a complex geometry. Hadronic acceleration is likely but the ultimate proof requires further observations. The evolution of the geometry along the orbit of $\eta$ Carinae provides a wealth of constrains that future observations and simulations will profit from.

\begin{acknowledgements}
We thank Ross Parkin for making the hydrodynamic simulation results available to us and Etienne Lyard for helping us interpret these files.
\end{acknowledgements}
\bibliographystyle{aa} % style aa.bst
\bibliography{29640} % your references Yourfile.bib

\end{document}